\documentclass[twocolumn,aps,floatfix,superscriptaddress,longbibliography]{revtex4-2}
\pdfoutput=1 
\usepackage{amsmath,amssymb,eucal,graphicx,float,epstopdf}
\usepackage{epsfig,bm}
\usepackage[utf8]{inputenc}

\usepackage[colorlinks=true, urlcolor=blue, anchorcolor=blue, citecolor=blue,filecolor=blue,linkcolor=blue,menucolor=blue,pagecolor=blue]{hyperref}

\begin{document}
\title{The Riviera model with egoistic settlers}

\author{P.~L.~Krapivsky}
\affiliation{Department of Physics, Boston University, Boston, Massachusetts 02215, USA}
\affiliation{Santa Fe Institute, Santa Fe, New Mexico 87501, USA}

\begin{abstract} 
The Riviera model mimics a densifying settlement along the coastline. In the lattice version, houses are built sequentially in empty sites with the constraint that every newly built house has at least one empty neighboring site. The distribution of clusters of adjacent houses does not obey a closed set of evolutionary equations, but the void-cluster-void distribution does. We compute the latter and extract the cluster distribution from it. In the jammed state, when all voids have length one, the cluster distribution takes a simple closed form and exhibits factorial decay with cluster size. For finite systems, we use a static approach that directly analyzes jammed states. If the coastline is a finite segment, we determine the statistics of the number of empty sites in the jammed state (the average, variance, and higher cumulants). We also study a continuum version in which houses are built along the line, subject to a minimum separation from at least one neighboring house.
\end{abstract}

\maketitle

\section{The Model}
\label{sec:RM}

The Riviera model is an idealized model of a settlement densifying along a beach. Houses are built sequentially subject to the constraint that each newly built house has access to sunlight from at least one side. More precisely \cite{Puljiz24}, the coastline is represented by a one-dimensional lattice. A house can be built at an empty site if either both neighboring sites are empty or one neighboring site is occupied but has another empty site on its other side. In the latter case, the newly built house continues to receive sunlight from that side. The selection of the building spots is random, and a house is built instantaneously, before the next attempt. Therefore, the process is sequential. The outcome is a jammed configuration locally looking like  
\begin{equation}
\label{jammed}
\cdots \circ\,\bullet\,\bullet\,\circ\,\circ\,\bullet\,\bullet\,\circ\,\bullet\,\circ\,\bullet\,\circ\,\bullet\,\circ\,\bullet\,\bullet\,\circ\,\bullet\,\circ\,\bullet\,\circ\cdots
\end{equation}
Here $\circ$ denotes an empty site, and $\bullet$ denotes a site with a house. The dynamical version of the Riviera model \cite{Puljiz24} remains unsolved despite significant effort \cite{JM23a,Pascal26}. One analytical advancement is the enumeration of jammed configurations with prescribed density on the coastline of prescribed length \cite{Puljiz24, JM23a}. The Riviera model arose as a one-dimensional version of the two-dimensional settlement model proposed in \cite{Puljiz22, Puljiz23}. 

There are many possible deformations of the Riviera model \cite{Puljiz24}. Instead of the lattice version with an infinite array of predetermined building sites, one can consider an off-lattice settlement, with houses built along a line with the constraint that each house (a segment of length $\ell$) remains separated by a certain minimal distance $w$ from at least one of its neighbors. This model also appears to be analytically intractable. The only exception is the extremal $w=0$ version that reduces to the so-called car-parking problem \cite{Renyi58}. Another possible deformation \cite{Puljiz25} concerns jammed configurations resistant to new types of agents (predators and altruists) operating after the settlement reaches a jammed state. 

Here, we analyze a model that can be interpreted as the Riviera model with egoistic settlers. Each settler builds a house if at least one neighboring site is empty, without considering that this neighbor may subsequently lose its only empty neighbor. We shall study this version and shortly refer to it as the Riviera model (RM). The RM belongs to a class of models introduced in \cite{Keller62}, and it is one of the simplest one-dimensional random sequential adsorption (RSA) processes (see \cite{Evans93,Talbot00,KRB} for review). 

The evolution of the void distribution in the RM was computed in \cite{Keller62}. We seek a more comprehensive characterization of jammed configurations and the approach toward them. In the jammed configuration of the RM, each void is a single site. A jammed configuration locally looks like  
\begin{equation}
\label{jammed-M}
\cdots \bullet\,\circ\,\bullet\,\bullet\,\bullet\,\bullet\,\circ\,\bullet\,\circ\,\bullet\,\circ\,\bullet\,\bullet\,\circ\,\bullet\,\bullet\,\bullet\,\circ\cdots
\end{equation}
Empty sites in the jammed configurations are isolated. The clusters of occupied sites have length 1 and 2 for the original Riviera model, and arbitrary length for the RM. Our first goal is to determine the distribution $\Pi_s$ of the clusters of length $s$, i.e., the density of patterns
\begin{equation}
\label{clusters}
\Pi_s = \text{Prob}\big[\!\circ \underbrace{\bullet \cdots \bullet}_{s} \circ\big]
\end{equation}
in the jammed state. The sum rule
\begin{equation}
\label{norm}
1 = \sum_{s\geq 1}(s+1)\Pi_s
\end{equation}
follows from the fact that, in the jammed state, the lattice is partitioned into clusters and single-site voids. The jamming density, i.e., the final fraction of occupied sites, is  
\begin{equation}
\label{empty}
\rho_\text{jam} = \sum_{s\geq 1}s\Pi_s = \frac{2}{3}
\end{equation}
The cluster-size distribution remains nontrivial in the jammed state: Clusters of all lengths, $s\geq 1$, are possible. One of our central results is the exact formula for the cluster-size distribution in the jammed state:
\begin{align}
\label{Pi-s}
\Pi_s = \frac{2^{s+1}\,s}{(s+4)(s+2)!} 
\end{align}

In the original Riviera model \cite{Puljiz24}, there are only clusters of size one and two. The voids in the jammed state also have size one or two, see \eqref{jammed}. Each void of size two separates clusters of size two. The densities of these short possible clusters of houses and voids are unknown. The original Riviera model \cite{Puljiz24} does not enjoy the shielding property:  The newly built house imposes the constraint that at least one of the two neighboring plots remains forever unbuilt. The shielding property is the hidden reason for the solvability of one-dimensional RSA models \cite{Evans93,Talbot00}. 

The RM enjoys the shielding property that makes the RM tractable and allows us to derive the densities \eqref{Pi-s} in the jammed state, and more generally the cluster densities $P_s(t)$ throughout the evolution. 

The distribution of voids in the RM is much easier to handle than the cluster-size distribution. The void-size distribution was established previously by Keller \cite{Keller62}. We outline the computation of the void-size distribution in Sec.~\ref{sec:evol} as it provides a simple setting in which to introduce the methods that we subsequently extend to the computation of cluster densities. Namely, in Sec.~\ref{sec:evol} we define the dynamics of the RM, write the governing equations for the densities 
\begin{equation}
\label{voids:seg}
V_n = \text{Prob}\big[\!\bullet \underbrace{\circ \cdots \circ}_{n} \bullet\big]
\end{equation}
and show that
\begin{equation}
\label{V:sol}
V_n(t) = \left(1-e^{-t}\right)^2\times 
\begin{cases}
\frac{1+2e^{-t}}{3} & n=1\\
e^{-nt}                   & n\geq 2
\end{cases}
\end{equation}
The fraction $e(t)$ of empty sites decays according to
\begin{equation}
\label{empty:t}
e(t)=\sum_{n\geq 1} n V_n(t) = \frac{1-e^{-3t}}{3}+e^{-2t}
\end{equation}
leading to \eqref{empty} in the jammed state ($t=\infty$). 

In Appendix~\ref{sec:one-sided}, we briefly analyze the RM on the semi-infinite lattice. We focus on the behavior near the boundary and show a remarkable phenomenon of {\em complete} saturation: The jamming density at any site apart from the leftmost site coincides with the bulk jamming density \eqref{empty}.

One can deform the dynamics by assigning different probabilities for building a house, depending on the number of surrounding empty sites. This class of models is also tractable, as demonstrated in Appendix~\ref{ap:gen}. The void distribution for this class of models, and also for slightly more general ones studied in \cite{Keller62}, is readily computable. 

In the jammed state, each void is a single site, so the final void distribution is trivial: $V_n(\infty) = \frac{1}{3}\delta_{n,1}$. In contrast, the cluster-size distribution remains non-trivial in the jammed state. The cluster-size distribution does not satisfy a closed set of equations. We therefore consider the more comprehensive void-cluster-void distribution, which satisfies a closed set of infinitely many linear ordinary differential equations. In Sec.~\ref{sec:clusters}, we solve these equations. From the solution, we extract the cluster-size distribution in the jammed state, \eqref{Pi-s}, as well as various dynamical predictions. 

The connected pair correlation function $C_j$ involves the occupancies of sites separated by distance $j$. In the jammed state, $C_j$ can be extracted from the cluster-size distribution \eqref{Pi-s} when $|j|\leq 3$. This is shown in Sec.~\ref{sec:PCF}. 

In Sec.~\ref{sec:segment}, we analyze finite systems. We determine the statistics of the number of empty sites (the average, the variance, and all higher cumulants) for the RM on a segment of length $L\gg 1$. We also compute the probabilities of reaching maximally sparse and maximally dense jammed states. Independently of the dynamics, one can count the number of jammed configurations, and also jammed configurations with a fixed number of houses. In Appendix~\ref{ap:entropy}, we determine the exact expressions for these quantities and extract the large $L$ behaviors from which we deduce the configurational entropy.

In Sec.~\ref{sec:line}, we analyze the continuum version of the RM where the line represents the coastline. This continuum counterpart of the lattice RM is akin to the R{\'e}nyi's car-parking model \cite{Renyi58}, which is a continuum counterpart of the lattice Flory model \cite{Flory39,Shepp}. Some results for the lattice RM can be extended to the continuum RM. However, the calculations are considerably more laborious. In Sec.~\ref{sec:line}, we limit ourselves to computing the void-size distribution from which we deduce the fraction of the uncovered line, the continuum analogs of \eqref{empty} and \eqref{empty:t}.  

In Sec.~\ref{sec:disc}, we discuss possible extensions of the same methods to computing the void-cluster-void-cluster-void distribution. In the jammed state, the non-vanishing  void-cluster-void-cluster-void densities are
\begin{equation}
\label{s1s2}
\Pi_{s_1,s_2} = \text{Prob}\big[\!\bullet \circ \underbrace{\bullet \cdots \bullet}_{s_1} \circ \underbrace{\bullet \cdots \bullet}_{s_2}  \circ\,\bullet\big]
\end{equation}
account for the correlation of sizes of neighboring clusters in the jammed state. These and higher-order densities $\Pi_{s_1,\ldots,s_k}$ are required for the computation of the connected pair correlation function with separation $j<2k$.

\section{Void-size distribution}
\label{sec:evol}

Even if we only want to describe the average characteristics of the jammed state, the experience with one-dimensional RSA processes suggests that it is convenient to treat the process dynamically. These processes are tractable when the evolution to the left of an occupied site is independent of the evolution to the right of the same occupied site. This is the shielding property \cite{Evans93, Talbot00}. The original Riviera model does not enjoy the shielding property since its evolution rule couples the neighboring sites of any occupied site. In contrast, the shielding property holds for the RM. The shielding property \cite{Evans93, Talbot00} has no analog in two and higher dimensions, and this is the chief reason why the RSA processes in $d\geq 2$ dimensions are intractable. 

We endow the RM with the following dynamics:
\begin{itemize}
\item Each site is chosen independently with the same rate that we set to one.
\item The house is built if the site is empty and at least one of its neighboring sites is empty. 
\end{itemize}
The RM dynamics is natural and convenient for analytical treatment. To simulate the process more efficiently, one can choose only empty sites. Even better is to choose a site from a (shrinking) list of accessible sites, i.e., empty sites with at least one neighboring empty site. Such algorithms correctly describe final jammed states. They also faithfully represent the dynamics used in the analytical work if, after each attempt, the time is increased by $1/N$ where $N$ is the number of empty sites in the case of the first algorithm and the number of accessible sites in the case of the second algorithm.

The densities of voids evolve according to an infinite set of linear ordinary differential equations (ODEs)
\begin{subequations}
\label{voids}
\begin{align}
\label{void-n}
\frac{dV_n}{dt} &=-n V_n +2\sum_{j\geq n+1}V_j, \qquad n\geq 2\\
\label{void-1}
\frac{dV_1}{dt} &=2\sum_{j\geq 2}V_j
\end{align}
\end{subequations}
The  loss term in \eqref{void-n} accounts for the houses built inside the voids of length $\geq 2$. The gain terms in Eqs.~\eqref{voids} describe the creation of voids when the house is built inside larger voids. Hereinafter, we typically suppress the dependence on time if there is no ambiguity. 

To solve Eqs.~\eqref{void-n} we employ an exponential ans\"{a}tze
\begin{equation}
\label{exp}
V_n = A a^{n} \qquad (n\geq 2)
\end{equation}
with yet unknown $A(t)$ and $a(t)$. Plugging \eqref{exp} into \eqref{void-n}  we reduce an infinite set of ODEs to a pair of ODEs:
\begin{subequations}
\label{aA:eq}
\begin{align}
\label{a:eq}
&\frac{da}{dt} = - a \\
\label{A:eq}
&A^{-1}\,\frac{dA}{dt} =  \frac{2a}{1-a}
\end{align}
\end{subequations}

The system is initially empty. Therefore
\begin{subequations}
\begin{equation}
\label{A:IC}
V_1(0)=0, \qquad A(0)=0
\end{equation}
Using relation $\sum_{n\geq 2} nV_n(0)=1-V_1(0)=1$ reflecting that the {\em entire} lattice was initially empty together with the exponential ans\"{a}tze \eqref{exp} we obtain 
\begin{equation*}
\label{Aa:IC}
\lim_{t\to 0} \frac{[2-a(t)]A(t)a^2(t)}{[1-a(t)]^2}=1
\end{equation*}
implying that 
\begin{equation}
\label{a:IC}
a(0)=1, \qquad \lim_{a\to 1} \frac{A}{(1-a)^2}=1
\end{equation}
\end{subequations}

Treating $A$ as a function of $a$ rather than time we recast \eqref{A:eq} into
\begin{equation}
\label{Aa:eq}
A^{-1}\,\frac{dA}{da} = -\frac{2}{1-a}
\end{equation}
from which $A = (1-a)^2$ where we additionally used \eqref{a:IC} to fix the amplitude. 

Using the exponential ans\"{a}tze \eqref{exp} we reduce the right-hand side (RHS) of \eqref{void-1} to $2Aa^2/(1-a)=2a^2(1-a)$.  We further divide \eqref{void-1} by \eqref{a:eq}, equivalently, treat $V_1$ as a function of $a$. We thus obtain
\begin{equation}
\label{V1a}
\frac{dV_1}{da} = 2a^2-2a
\end{equation}
which we integrate subject to $V_1(a=1)=0$ to find
\begin{subequations}
\label{VV-a}
\begin{equation}
\label{V1-a}
V_1=\frac{1+2a}{3}\,(1-a)^2
\end{equation}
The densities of longer voids are
\begin{equation}
\label{exp-a}
V_n = (1-a)^2 a^{n}, \qquad n\geq 2
\end{equation}
\end{subequations}

We now return to the original time variable. Integrating \eqref{a:eq} subject to $a(0)=1$ yields $a = e^{-t}$, so Eqs.~\eqref{VV-a} lead to the announced densities \eqref{V:sol}. 

Instead of voids \eqref{voids:seg}, one can consider segments with consecutive empty sites with unspecified bordering sites:
\begin{equation}
\label{voids:seg+}
W_n = \text{Prob}\big[\!\underbrace{\circ \cdots \circ}_{n}\big]
\end{equation}
These densities evolve according to equations
\begin{subequations}
\label{voids+}
\begin{align}
\label{void-n+}
\frac{dW_n}{dt} &=-n W_n, \qquad n\geq 2\\
\label{void-1+}
\frac{dW_1}{dt} &=-2W_2+W_3
\end{align}
\end{subequations}
When $n\geq 2$, the evolution of $W_n$ is independent of the state of the bordering sites explaining \eqref{void-n+}. The evolution of $W_1$ depends on the bordering sites, viz., the process $\circ\to\bullet$ is possible when an empty site is surrounded by at most one occupied site: $\circ \circ \circ$, $\circ \circ \bullet$, or $\bullet \circ \circ$. The corresponding rates are $W_3$, $W_2-W_3$, $W_2-W_3$, i.e., the overall rate $2W_2-W_3$, leading to \eqref{void-1+}. 

Equations \eqref{voids+} are simpler than equations \eqref{voids} describing the evolution of the densities of voids. Indeed, Eqs.~\eqref{void-n} are not even recursive, while Eqs.~\eqref{void-n+} are decoupled and trivially solvable. The solvability of Eqs.~\eqref{void-n} rests on the exponential ans\"{a}tze. However, this method is inapplicable if the initial conditions are incompatible with the exponential ans\"{a}tze. The initial conditions, $W_n(0)=1$ for all $n\geq 1$, for Eqs.~\eqref{voids+} are also simpler than the subtle initial conditions \eqref{A:IC}  and \eqref{a:IC} for \eqref{aA:eq}. 

Solving \eqref{void-n+} and then \eqref{void-1+} we obtain
\begin{subequations}
\label{Wn:sol}
\begin{align}
\label{void-n+:sol}
&W_n = e^{-nt}\,, \qquad n\geq 2\\
\label{void-1+:sol}
&W_1 = \frac{1-e^{-3t}}{3}+e^{-2t}
\end{align}
\end{subequations}
The general relation 
\begin{equation}
\label{Vn:W}
V_n = W_n - 2 W_{n+1}+W_{n+2}
\end{equation}
agrees with \eqref{Wn:sol} and \eqref{V:sol}.

\section{Cluster-size distribution}
\label{sec:clusters}

To determine the distribution of clusters of occupied sites in the jammed state given by Eq.~\eqref{clusters}, one can try to write evolution equations for the same densities $P_s(t)$ at time $t$. However, these densities do not obey a closed set of equations. One can consider clusters with more detailed description of the boundary sites:
\begin{subequations}
\begin{align}
\label{X:seg}
& \circ \circ \underbrace{\bullet \cdots \bullet}_{s} \circ \circ  \qquad X_s\\
\label{Y:seg}
& \bullet \circ\, \underbrace{\bullet \cdots \bullet}_{s} \circ \circ  \qquad Y_s\\
\label{Z:seg}
& \bullet \circ\, \underbrace{\bullet \cdots \bullet}_{s} \circ \bullet  \qquad Z_s
\end{align}
\end{subequations}
The densities $X_s, Y_s, Z_s$ satisfy the evolution equations 
\begin{equation*}
\begin{split}
& \frac{d X_s}{dt} = - 4 X_s + (\cdots)\\
& \frac{d Y_s}{dt} = - 2 Y_s + (\cdots)\\
& \frac{d Z_s}{dt} =  2 Y_s + (\cdots)
\end{split}
\end{equation*}
The displayed terms are exact, but one cannot express the omitted gain terms $(\cdots)$ via the quantities $X, Y,Z$. For instance 
\begin{equation*}
\star \circ \circ \underbrace{\bullet \cdots \bullet}_{s-1} \circ \circ \longrightarrow \star \circ \underbrace{\bullet \cdots \bullet}_{s} \circ \circ 
\end{equation*}
contributes to $X_s$ if $\star=\circ$ and to $Y_s$ if $\star=\bullet$.

To determine $P_s$, we need a better set of densities than $X_s, Y_s, Z_s$. Clusters appear in several RSA models. For instance, clusters are formed in the one-dimensional monolayer ballistic deposition model \cite{Talbot92,Talbot93} that generalizes ballistic deposition and the continuum version \cite{Renyi58} of the RSA. The void-cluster-void distribution was used in the computation of the cluster densities in Ref.~\cite{Talbot93}. Below we apply the same idea to the RM. We thus consider the patterns
\begin{equation}
\label{void-cluster-void}
\bullet \underbrace{\circ \cdots \circ}_{m}\underbrace{\bullet \cdots \bullet}_{s} \underbrace{\circ \cdots \circ}_{n}\bullet 
\end{equation}
with $m,s, n\geq 1$ and write the evolution equations for the densities $P_s(m,n; t)$ of such patterns: 
\begin{eqnarray}
\label{P-mns}
\frac{dP_s(m,n)}{dt} &=& - [m+n-\delta_{m,1}-\delta_{n,1}] P_s(m,n) \nonumber \\
&+&\sum_{m'\geq m+1}P_s(m',n)+\sum_{n'\geq n+1}P_s(m,n')\nonumber \\
&+&[1-\delta_{s,1}][P_{s-1}(m+1,n)+ P_{s-1}(m,n+1)]\nonumber \\
&+& \delta_{s,1}V_{m+1+n}
\end{eqnarray}
The loss terms describe the appearance of the houses in voids of width $m\geq 2$ and $n\geq 2$. The specificity of the voids of the minimal width is reflected by Kronecker delta symbols $\delta_{m,1}$ and $\delta_{n,1}$ in the loss terms. When $m,n\geq 2$, the loss terms in Eqs.~\eqref{P-mns} become $(m+n)P_s(m,n; t)$, suggesting seeking the solution in the form
\begin{equation}
\label{Qs:def}
P_s(m,n; t) = e^{-(m+n)t} Q_s(t)\qquad (m,n\geq 2)
\end{equation}
This ans\"{a}tze reduces Eqs.~\eqref{P-mns} with $m\geq 2$ and $n\geq 2$ to  
\begin{subequations}
\label{Q-eq}
\begin{align}
\label{Q1}
\frac{d Q_1}{dt} &= \frac{2}{e^t-1}\,Q_1+e^{-t}\left(1-e^{-t}\right)^2  &  s=1\\
\label{Qs}
\frac{d Q_s}{dt} &= \frac{2}{e^t-1}\,Q_s + 2e^{-t}Q_{s-1}                 &  s\geq 2
\end{align}
\end{subequations}
We thus have linear ODEs for $Q_s(t)$ with $s\geq 1$ rather than linear hierarchical ODEs for the three-parameter $P_s(m,n; t)$ with $m,n\geq 2$ and $s\geq 1$. This property is a technical simplification. The crucial simplifying feature is the recursive nature of Eqs.~\eqref{Q-eq}. The evolution equations \eqref{P-mns} are not recursive.

Instead of the original time variable it proves convenient to use an auxiliary time variable
\begin{equation}
\label{tau:def}
\tau = 1-e^{-t}
\end{equation}
In terms of this new time variable, Eqs.~\eqref{Q-eq} become
\begin{subequations}
\begin{align}
\label{Q1:tau}
\frac{d Q_1}{d\tau} &= \frac{2Q_1}{\tau}+\tau^2                   &  s=1\\
\label{Qs:tau}
\frac{d Q_s}{d\tau} &= \frac{2Q_s}{\tau} + 2Q_{s-1}             &  s\geq 2
\end{align}
\end{subequations}
Solving \eqref{Q1:tau} subject to $Q_1(0)=0$ gives $Q_1=\tau^3$. We then recurrently solve Eqs.~\eqref{Qs:tau} subject to $Q_s(0)=0$ and find
\begin{equation}
\label{Qs:sol}
Q_s(\tau) = \frac{2^{s-1}}{s!}\,\tau^{s+2}
\end{equation}
applicable to all $s\geq 1$. 

The following analysis depends on whether two or one of the bordering voids have minimal width. In the former case when one of the two voids has the minimal width, e.g., $m=1$, the loss terms become $nP_s(1,n; t)$, suggesting seeking the solution in the form 
\begin{equation}
\label{Rs:def}
P_s(1,n; t) = e^{-nt} R_s(t)\qquad (n\geq 2)
\end{equation}
This ans\"{a}tze greatly simplifies Eqs.~\eqref{P-mns} with $m=1$ and $n\geq 2$ reducing them to
\begin{subequations}
\begin{equation}
\label{R1:tau}
\frac{d R_1}{d\tau} = \frac{R_1}{\tau} + (1-\tau)\,\frac{Q_1}{\tau}+\tau^2(1-\tau)       
\end{equation}
and
\begin{align}
\label{Rs:tau}
\frac{d R_s}{d\tau} = \frac{R_s}{\tau} +R_{s-1} +(1-\tau)\left[\frac{Q_s}{\tau} + Q_{s-1}\right] 
\end{align}
\end{subequations}
for $s\geq 2$. Since $Q_1=\tau^3$, Eq.~\eqref{R1:tau} becomes 
\begin{subequations}
\label{R-eq}
\begin{align}
\label{R1:eq}
\frac{d R_1}{d\tau}  = \frac{R_1}{\tau} +2\tau^2(1-\tau)       
\end{align}
When $s\geq 2$, we use \eqref{Qs:sol} and re-write \eqref{Rs:tau} as
\begin{align}
\label{Rs:eq}
\frac{d R_s}{d\tau} = \frac{R_s}{\tau} + R_{s-1} +  \frac{2^{s-2}(s+2)}{s!}\,\tau^{s+1}(1-\tau)
\end{align}
\end{subequations}
Solving \eqref{R1:eq} subject to $R_1(0)=0$ gives $R_1=\tau^3-\frac{2}{3}\tau^4$. We then recursively solve \eqref{Rs:eq} subject to $R_s(0)=0$ and find $R_2=\tau^4-\frac{2}{3}\tau^5$, $R_3=\frac{2}{3}\tau^5-\frac{7}{15}\tau^6$,  etc. from which we conclude that $R_s=A_s \tau^{s+2}+B_s \tau^{s+3}$. Substituting this ans\"{a}tze into Eqs.~\eqref{Rs:eq} we fix the amplitudes $A_s$ and $B_s$ to give
\begin{equation}
\label{Rs:sol}
R_s(\tau)=\frac{2^{s-1}}{s!}\,\tau^{s+2}- \frac{(s^2+s+2)2^{s-1}}{(s+2)!}\, \tau^{s+3}
\end{equation}
Equation \eqref{Rs:sol} is applicable to all $s\geq 1$. 

\begin{figure}
\centering
\includegraphics[width=8.12cm]{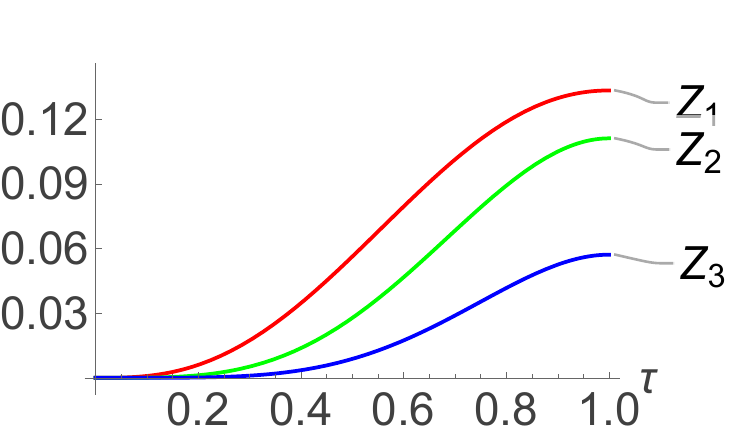}
\caption{The densities $Z_1, Z_2, Z_3$ are growing functions of time. This is a common feature of all densities $Z_s$. The final jammed densities are  $\Pi_1=\frac{2}{15}, \Pi_2=\frac{1}{9}, \Pi_3=\frac{2}{35}$.}
\label{Fig:Z123}
\end{figure}

When both voids have minimal width, $m=n=1$, the corresponding density $P_s(1,1)\equiv Z_s$, see \eqref{Z:seg}, obeys
\begin{subequations}
\label{Z:eq}
\begin{align}
\label{Z1:eq}
\frac{d Z_1}{d\tau} &= 2(1-\tau)\,\frac{R_1}{\tau}  + \tau^2(1-\tau)^2 \\
\label{Zs:eq}
\frac{d Z_s}{d\tau} &= 2(1-\tau)\left[\frac{R_s}{\tau}  + R_{s-1}\right] \quad (s\geq 2)
\end{align}
\end{subequations}
Inserting \eqref{Rs:sol} into \eqref{Z:eq} and integrating we obtain
\begin{eqnarray}
\label{Zs:sol}
Z_s &=& 2^{s-1}\frac{s^3 + 7 s^2 + 14 s + 8}{(s+4)(s+2)!}\,\tau^{s+2} -2^s\,\frac{s^2+s+2}{(s+2)!}\,\tau^{s+3}  \nonumber \\
&+& 2^{s-1}\,\frac{s^3+3s^2+2s+8}{(s+4)(s+2)!}\,\tau^{s+4}
\end{eqnarray}
applicable for all $s\geq 1$. The first three densities are plotted in Fig.~\ref{Fig:Z123}.

Equations \eqref{Qs:sol}, \eqref{Rs:sol}, \eqref{Zs:sol} together with \eqref{Qs:def} and \eqref{Rs:def} defining $Q_s$ and $R_s$ constitute the exact solution for the void-cluster-void size distribution. The chief reason for the solvability of Eqs.~\eqref{P-mns} is the applicability of the exponential substitutions \eqref{Qs:def} and \eqref{Rs:def} reducing Eqs.~\eqref{P-mns} to much simpler (but still infinite) sets of linear ODEs, Eqs.~\eqref{Q-eq} and \eqref{R-eq}, which are {\em recursive} and hence solvable. The exponential substitutions are applicable in our case when the initial state is empty; they are not valid for an arbitrary initial condition. 

\begin{figure}
\centering
\includegraphics[width=8.12cm]{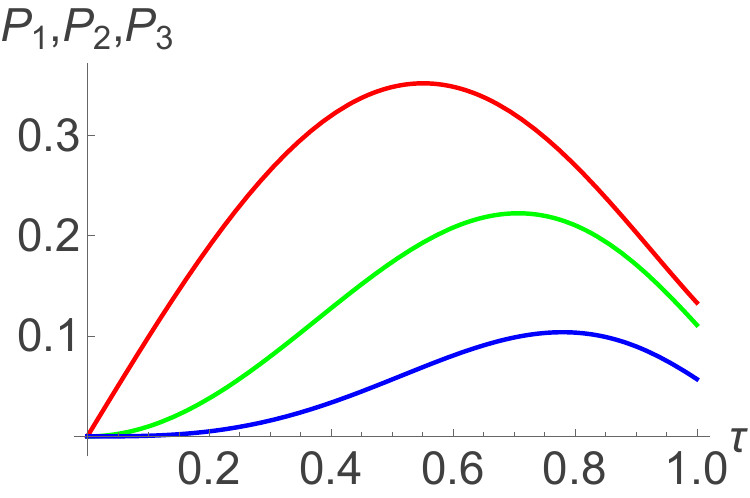}
\caption{The cluster densities $P_1, P_2, P_3$ (top to bottom) versus the auxiliary time $\tau=1-e^{-t}$. Cluster densities $P_s(\tau)$ are unimodal distributions with a single peak at $\tau_s$ given by \eqref{tau-s}; $\tau_1=\sqrt{\frac{1}{7} \left(6-\sqrt{15}\right)}\approx 0.551234, \,\,\tau_2=2^{-1/2}\approx 0.707107, \tau_3=\sqrt{\frac{1}{34} \left(35-\sqrt{205}\right)}\approx 0.779935$, etc.}
\label{Fig:P123}
\end{figure}

The cluster-size distribution $P_s$ can be expressed via the void-cluster-void distribution
\begin{equation}
\label{P:PPZ}
P_s = \sum_{m,n\geq 2}P_s(m,n)+ 2\sum_{n\geq 2}P_s(1,n)+Z_s
\end{equation}
Using \eqref{Qs:def} and \eqref{Rs:def} we simplify \eqref{P:PPZ} to
\begin{equation}
\label{P:QRZ}
P_s = \left(\frac{1-\tau}{\tau}\right)^2 Q_s+ 2\, \frac{1-\tau}{\tau}\, R_s +Z_s
\end{equation}
or equivalently
\begin{equation}
\label{Ps:sol}
P_s = \frac{2^{s-1}\tau^{s}}{s!}\left[1-A_s\tau^2+B_s\tau^4\right]
\end{equation}
where 
\begin{equation*}
A_s=2\,\frac{s^2+s+2}{(s+2)(s+1)}\,, \quad B_s = \frac{s^3+3s^2+2s+8}{(s+4)(s+2)(s+1)}
\end{equation*}
The densities $P_s(\tau)$ are unimodal with a single peak at
\begin{equation}
\label{tau-s}
\tau_s = \sqrt{\frac{s^2+s}{s^2+s+2+2 \sqrt{\frac{s^3+s^2+s+2}{s+2}}}}
\end{equation}
The first three densities are plotted in Fig.~\ref{Fig:P123}. 

In the jammed state, $\Pi_s=P_s(\tau=1)=Z_s(\tau=1)$. Specializing \eqref{Zs:sol} to $\tau=1$ we arrive at the announced cluster-size distribution \eqref{Pi-s} in the jammed state.

\section{Pair Correlation Function}
\label{sec:PCF}

The connected equal-time pair correlation function is defined via 
\begin{equation*}
\label{Cj:def}
C_j(t) = \langle\!\langle O_0(t)O_i(t)\rangle\!\rangle= \langle O_0(t) O_j(t)\rangle -  \langle O_0(t)\rangle  \langle O_j(t)\rangle
\end{equation*}
Here $O_j$ is the indicator function for the occupancy of site $j$ (occupancy, in short): $O_j=0$ if site $j$ is empty and $O_j=1$ otherwise. The average occupancy is spatially homogeneous, $\langle O_j(t)\rangle=1-e(t)$ for all $j\in \mathbb{Z}$, with $e(t)$ given by Eq.~\eqref{empty:t}. The computations of $C_j(t)$ are laborious even in the simplest one-dimensional RSA processes \cite{Hilhorst,Hemmer,Pascal94}; see \cite{Evans93,Talbot00,KRB} for review. Performing such computations in the RM is left for the future. Here, we determine $C_j$ with $|j|\leq 3$ directly from the cluster-size distribution. 

In what follows, we focus on the jammed state. Hereinafter, we write $C_j$ instead of $C_j(\infty)$. The sum of $C_j$ over $j\in \mathbb{Z}$ is related to the variance of the number of houses (equivalently empty sites) by the simple relation
\begin{equation}
\label{Cj:sum-var}
\sum_{j=-\infty}^\infty C_j = \lim_{L\to\infty}L^{-1}  \langle\!\langle N^2\rangle\! \rangle
\end{equation}
following from the definitions of the connected pair correlation function and the variance \cite{KRB}. In the next section, we compute the variance in the jammed state, show that it increases linearly in $L$ with amplitude $\frac{2}{45}$. Therefore
\begin{equation}
\label{Cj:sum}
\sum_{j=-\infty}^\infty C_j = \frac{2}{45}
\end{equation}

Using $C_0=\langle O^2\rangle-\langle O\rangle^2\equiv \langle O\rangle-\langle O\rangle^2$ and the average occupancy $\langle O\rangle=\frac{2}{3}$ in the jammed state we obtain 
\begin{equation}
\label{C0}
C_0=\frac{2}{9}
\end{equation}
(The identity $O^2\equiv O$ is valid for occupancy, and generally for any binary $\{0,1\}$ variable.) 

In addition to $C_0$, we computed  
\begin{equation}
\label{C123}
C_1 = -\frac{1}{9}\,, \qquad C_2 = \frac{1}{45}\,, \qquad C_3 = 0
\end{equation}
in the jammed state. The derivations of \eqref{C123}  $C_j$ rely on expressing $\langle O_0 O_j\rangle$ with $j\leq 3$ through the cluster-size distribution $\Pi_s$ in the jammed state. First, we use the identity 
\begin{equation}
\label{n01}
\langle O_0 O_1\rangle = \sum_{s\geq 1}(s-1)\Pi_s
\end{equation}
and the sum rules \eqref{norm} and \eqref{empty} to find $\langle O_0 O_1\rangle=\frac{1}{3}$ which in conjunction with  $\langle O\rangle=\frac{2}{3}$ yield $C_1 = -\frac{1}{9}$. 

Similarly to \eqref{n01} we find
\begin{equation}
\label{n02}
\langle O_0 O_2\rangle = \sum_{s\geq 2}(s-2)\Pi_s+\sum_{s\geq 1}\Pi_s 
\end{equation}
The first term on the right-hand side (RHS) of Eq.~\eqref{n02} accounts for the contribution to $\langle O_0 O_2\rangle$ from the sites in the same cluster; the second term is the contribution from the right-most site of any cluster and the left-most site of the neighboring cluster on the right. Simplifying the sums in \eqref{n02} we obtain 
\begin{equation*}
\langle O_0 O_2\rangle = \Pi_1+\sum_{s\geq 1}(s-1)\Pi_s = \frac{2}{15}+\frac{1}{3} = \frac{7}{15}
\end{equation*}
from which we deduce $C_2 = \langle O_0 O_2\rangle - \langle n\rangle^2 = \frac{1}{45}$. 

Similarly to \eqref{n02} we get
\begin{equation}
\label{n03}
\langle O_0 O_3\rangle = \sum_{s\geq 3}(s-3)\Pi_s+\sum_{s\geq 2}\Pi_s + \sum_{s\geq 2}\Pi_s 
\end{equation}
The first term on the RHS accounts for the contribution to $\langle O_0 O_3\rangle$ from the sites in the same cluster. The second term is the contribution from the left-most site of any cluster and the penultimate site of the neighboring cluster on the left, i.e., the pattern $\bullet\bullet\circ\,\bullet$, while the third term is the contribution from the symmetric pattern $\bullet\circ\bullet\,\bullet$. Simplifying the sums on the RHS of \eqref{n03} we obtain $\Pi_2+\sum_{s\geq 1}(s-1)\Pi_s$. Recalling the sum rules \eqref{norm}, \eqref{empty} and $\Pi_2=\frac{1}{9}$ we obtain $\langle O_0 O_3\rangle = \frac{4}{9}$ leading to 
$C_3=0$. 

The cluster-size distribution \eqref{Pi-s} is insufficient to determine $C_j$ for $j\geq 4$. For instance, to determine $C_4$, we need to compute
\begin{equation}
\label{n04}
\langle O_0 O_4\rangle = \sum_{s\geq 4}(s-4)\Pi_s+2\sum_{s\geq 3}\Pi_s + \text{Prob}[\bullet\bullet\circ\bullet\,\bullet]
\end{equation}
The terms on the RHS are analogous to the terms on the RHS of \eqref{n03}, apart from the last term depending on the correlations between neighboring clusters. Computing such correlations should be feasible, as we argue in Sec.~\ref{sec:disc}. However, such a computation appears significantly more laborious than the computation of the cluster-size distribution presented in Sec.~\ref{sec:clusters}.

We now outline an alternative and simpler derivation of \eqref{C0}--\eqref{C123} and the challenge of computing $C_j$ for $j\geq 4$. The trick is to rely on the indicator function for empty sites: $E_j=1$ if site $j$ is empty and $E_j=0$ otherwise. We define the associated connected pair correlation function $C_j=\langle\!\langle E_0 E_i\rangle\!\rangle= \langle E_0 E_j\rangle -  \langle E_0\rangle \langle E_0\rangle$ and use $E_j=1-O_j$ to check that this definition gives the same connected pair correlation function. We have $E_0^2=E_0$ again. Furthermore, $E_0 E_1=0$ in the jammed state. Therefore $\langle E_0^2\rangle = \frac{1}{3}$ and $\langle E_0 E_1\rangle = 0$ in the jammed state from which we recover $C_0=\frac{2}{9}$ and $C_1=-\frac{1}{9}$. We also have $\langle E_0 E_2\rangle = \Pi_1=\frac{2}{15}$ from which we recover $C_2=\frac{1}{45}$. Similarly, $\langle E_0 E_3\rangle = \Pi_2=\frac{1}{9}$ providing a simpler derivation of $C_3=0$ than the previous derivation. 

For $j\geq 4$, the simple relation $\langle E_0 E_j\rangle = \Pi_{j-1}$ is no longer valid. When $j=4$, we get
\begin{equation}
\label{e04}
\langle E_0 E_4\rangle = \Pi_3 + \Pi_{1,1}, \quad \Pi_{1,1}=\text{Prob}[\circ\bullet\circ\bullet\circ]
\end{equation}
This equation is simpler than \eqref{n04} but it does not give us $\langle E_0 E_4\rangle$ since $\Pi_{1,1}$ depending on the correlations between neighboring clusters of minimal length is unknown. The quantity $P_{1,1}(t)$ does not satisfy a closed evolution equation, and it can be established only after computing a complicated void-cluster-void-cluster-void distribution, see Sec.~\ref{sec:disc}. Similarly 
\begin{equation}
\label{e56}
\begin{split}
\langle E_0 E_5\rangle &= \Pi_4 + 2\Pi_{1,2}\\
\langle E_0 E_6\rangle & = \Pi_3 + 2\Pi_{1,3} + \Pi_{2,2} + \Pi_{1,1,1}
\end{split}
\end{equation}
involving unknown correlation functions
\begin{equation*}
\begin{split}
\Pi_{1,2} & =\text{Prob}[\circ\bullet\circ\bullet\bullet\,\circ]\\
\Pi_{1,3}& = \text{Prob}[\circ\bullet\circ\bullet\bullet\bullet\,\circ] \\
\Pi_{2,2}& = \text{Prob}[\circ\bullet\bullet\circ\bullet\bullet\,\circ] \\
\Pi_{1,1,1}& = \text{Prob}[\circ\bullet\circ\bullet\circ\bullet\,\circ] 
\end{split}
\end{equation*}

Using the symmetry relation, $C_{-j}=C_j$, together with \eqref{C0} and \eqref{C123} we transform \eqref{Cj:sum} into
\begin{equation}
\label{C3-sum}
\sum_{j\geq 4} C_j = 0
\end{equation}

The simplest conjecture compatible with $C_3=0$ and \eqref{C3-sum} is $C_j = 0$ for $j\geq 3$. If this holds, the system exhibits complete decorrelation beyond distance two. Whether this conjecture is correct requires the higher-order cluster correlations discussed in Sec.~\ref{sec:disc}. 

The vanishing of the pair correlation function beyond a finite distance is known as complete decorrelation. This remarkably rare phenomenon occurs in a very few systems \cite{Torquato16,Rahul23}. Complete decorrelation is more common in high dimensions \cite{TS06a,TS06b,Torquato08a, Cohn18}. On general grounds, complete decorrelation in a one-dimensional RM would be unexpected. However, our version of the RM in which an empty site surrounded by two or one empty sites are equally attractive seems rather special. In the settings with boundaries, viz., on segments and semi-infinite lattice, we established complete saturation of the jamming density one site away from the boundaries, cf. \eqref{jam:segment} and \eqref{jam:semi}. This saturation property is reminiscent of complete decorrelation.

\section{Finite Systems}
\label{sec:segment}

Here we consider the RM on a segment of length $L$. We should precisely define the rules of building a house in boundary sites. The specific choice 
\begin{equation}
\label{seg}
\blacktriangleleft \underbrace{\circ \cdots \circ}_{L} \blacktriangleright
\end{equation}
implies that the boundary sites do not receive the sunlight from the outside. This convention allows us to apply the rules of the RM defined in Sec.~\ref{sec:evol} to all sites. 

We focus on the jammed states that vary from realization to realization. In a finite system, one can explore fluctuations in the total number of empty sites, the probabilities of maximally dense and maximally sparse jammed states, etc. For concreteness, we consider the total number $N$ of empty sites in the final state; the total number of houses $H$ is a complementary random quantity, $H+N=L$, so it suffices to study one of these two random quantities. 

The maximal and minimal possible numbers of empty sites are 
\begin{equation}
\label{min-max}
N_\text{min} = 1, \qquad N_\text{max} = \left\lfloor \frac{L+1}{2}\right\rfloor
\end{equation}

To appreciate these extremal values we note that the maximally sparse jammed configuration is 
\begin{subequations}
\label{sparse}
\begin{equation}
\label{sparse:odd}
\blacktriangleleft\underbrace{ \bullet \circ \bullet \circ \bullet \cdots \circ \bullet}_{L}\blacktriangleright
\end{equation}
when $L$ is odd. When $L$ is even, there are two maximally sparse jammed configurations 
\begin{equation}
\label{sparse:even}
\blacktriangleleft\underbrace{\circ \bullet \circ \bullet \cdots \circ \bullet}_{L}\blacktriangleright    \quad\text{and}\quad  
\blacktriangleleft\underbrace{\bullet \circ \bullet \cdots \circ \bullet\, \circ}_{L}\blacktriangleright
\end{equation}
\end{subequations}
The number of empty sites is $N=\left\lfloor \frac{L+1}{2}\right\rfloor$, and this is the maximal possible number. 

The maximally dense jammed configurations 
\begin{equation}
\label{dense}
\blacktriangleleft\underbrace{\bullet\cdots \bullet \circ \bullet \cdots \bullet}_{L}\blacktriangleright 
\end{equation}
have a single empty site, $N=1$, located at an arbitrary position. There are $L$ maximally dense jammed configurations.

\subsection{The average number of empty sites}
\label{subsec:av}

We now present the computation of the average number $A_L=\langle N\rangle$ of empty sites in a segment of length $L$ in the final (jammed) state. We employ an approach essentially developed by Flory \cite{Flory39}; see also a textbook \cite{KRB}. For small $L$, one computes $A_L$ by hand and finds 
\begin{equation}
A_1=1,\quad A_2=1, \quad A_3 = \frac{4}{3}, \quad A_4 = \frac{5}{3}
\end{equation}

Suppose the first house is built at site $k$. Houses in segments of lengths $k-1$ and $L-k$ on the left and right of the first house are built {\em independently}. This crucial shielding property \cite{Evans93,Talbot00} makes the model tractable. The average number of empty sites satisfies the recurrence
\begin{equation}
\label{EL:rec}
A_L = \frac{1}{L}\sum_{k=1}^L \big(A_{k-1}+A_{L-k}\big)
\end{equation}
applicable for all $L\geq 2$ if we set $A_0=0$. Using the generating function 
\begin{equation}
\label{Ex:def}
\mathcal{A}(x) = \sum_{L\geq 1} A_{L}\, x^L
\end{equation}
we convert the recurrence \eqref{EL:rec} into a differential equation
\begin{equation}
\frac{d\mathcal{A}}{dx} = \frac{2}{1-x}\,\mathcal{A}+1
\end{equation}
which we solve subject to $\mathcal{A}(0)=0$ and find
\begin{equation}
\label{Ex:sol}
\mathcal{A}(x) = \frac{(1-x)^{-2}+1-x}{3}
\end{equation}
Expanding the generating function we arrive at
\begin{equation}
\label{EL:sol}
A_L = \frac{L+1}{3}\qquad \text{for}\quad L\geq 2
\end{equation}

\subsection{Full counting statistics}
\label{subsec:FCS}

The total number $N$ of empty sites in the final state fluctuates from realization to realization. The full statistics is determined by the probability $P(N,L)$ to have $N$ empty sites in a jammed state. It proves convenient to deal with the generating function 
\begin{equation}
\label{F:def}
F(\lambda, L) \equiv \langle e^{\lambda N}\rangle = \sum_N e^{\lambda N} P(N,L)
\end{equation}
encoding cumulants $\langle\!\langle N^n\rangle\!\rangle$. Indeed, the standard relation
\begin{equation}
\ln \langle e^{\lambda N}\rangle = \sum_{n\geq 1} \frac{\lambda^n}{n!}\, \langle\!\langle N^n\rangle\!\rangle
\end{equation}
gives the cumulants: The average $\langle\!\langle N\rangle\!\rangle = \langle N\rangle$, the variance $\langle\!\langle N^2\rangle\!\rangle = \langle N^2\rangle - \langle N\rangle^2$, and all higher cumulants. The function $F(\lambda, L) \equiv \langle e^{\lambda N}\rangle$ grows exponentially with $L$. Hence, we define the cumulant generating function 
\begin{equation}
\label{U:def}
U(\lambda) = \lim_{L\to\infty} L^{-1} \ln F(\lambda, L)
\end{equation}
encapsulating all cumulants: 
\begin{equation}
\label{U-cum}
U(\lambda)=\sum_{n\geq 1} \frac{\lambda^n}{n!}\, U_n, \qquad \langle\!\langle N^n\rangle\!\rangle = L U_n
\end{equation}

The function $F(\lambda, L)$ satisfies the recurrence 
\begin{equation}
\label{FL:rec}
F(\lambda, L) = \frac{1}{L}\sum_{k=1}^L F(\lambda, k-1) F(\lambda, L-k)
\end{equation}
which is a multiplicative analog of the recurrence \eqref{EL:rec} for the average. The recurrence \eqref{FL:rec} is derived similarly to \eqref{EL:rec}. Indeed, suppose the first house is built at site $k$. The houses on the left and right of site $k$ are built independently. If $N_-$ and $N_+$ are the final numbers of empty sites on the left and right, the total number of empty sites is $N= N_-  +N_+$, so $e^{\lambda N}=e^{\lambda N_-}e^{\lambda N_+}$. Performing the averaging, summing over all $k$, and recalling that site $k$ is chosen with probability $L^{-1}$, we obtain Eq.~\eqref{FL:rec}. 

Since $N=1$ when $L=1$, we have
\begin{subequations}
\begin{equation}
\label{F:1}
F(\lambda, 1)=e^\lambda
\end{equation}
The recurrence \eqref{FL:rec} applies for all $L\geq 2$ if we set 
\begin{equation}
\label{F:0}
F(\lambda, 0)=1
\end{equation}
\end{subequations}
For instance, $N=1$ when $L=2$, so $F(\lambda, 2)=e^\lambda$, and this is indeed recovered after specializing \eqref{FL:rec} to $L=2$ and using \eqref{F:1} and \eqref{F:0}. 

The shielding property was crucial for deriving the recurrence \eqref{EL:rec}. This recurrence is the convolution suggesting to introduce the generating function
\begin{equation}
\label{Phi:def}
\mathcal{F}(\lambda, x) = \sum_{L\geq 0}  F(\lambda, L)\, x^L
\end{equation}
We recast the recurrence \eqref{FL:rec} into an equation for the generating function by multiplying \eqref{FL:rec} by $Lx^{L-1}$ and summing over all $L\geq 2$. The left-hand side turns into
\begin{equation*}
\sum_{L\geq 2}  L F(\lambda, L)\, x^{L-1} = \frac{\partial \mathcal{F}}{\partial x} - e^\lambda
\end{equation*}
The right-hand side becomes
\begin{equation*}
\sum_{L\geq 2} \sum_{k=1}^L F(\lambda, k-1) F(\lambda, L-k)\, x^{L-1}=\mathcal{F}^2 -1
\end{equation*}
Therefore
\begin{equation}
\label{Phi:eq}
\frac{\partial \mathcal{F}(\lambda, x)}{\partial x} = [\mathcal{F}(\lambda, x)]^2 + e^\lambda - 1
\end{equation}

\begin{figure}[ht]
\begin{center}
\includegraphics[width=0.48\textwidth]{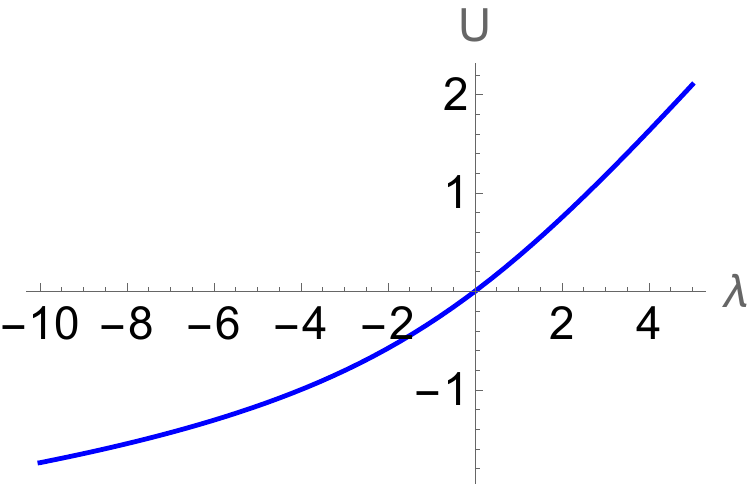}
\caption{The plot of the cumulant generating function $U(\lambda)$ given by Eq.~\eqref{U:sol}. The expansion of $U(\lambda)$ at $\lambda=0$ yields the cumulants.}
\label{Fig:U}
  \end{center}
\end{figure}

Solving the Riccati equation \eqref{Phi:eq} subject to the `initial' condition $\mathcal{F}(\lambda,0)=1$ implied by \eqref{F:0} gives
\begin{equation}
\label{Phi:sol}
\mathcal{F}(\lambda, x) = \frac{(e^\lambda-1)\tan\!\big(x\sqrt{e^\lambda -1}\big)+\sqrt{e^\lambda-1}}{\sqrt{e^\lambda-1}-\tan\!\big(x\sqrt{e^\lambda -1}\big)}
\end{equation}
The generating function \eqref{Phi:sol} has a simple pole at 
\begin{equation}
x_*(\lambda) = \frac{\arctan\sqrt{e^\lambda-1}}{\sqrt{e^\lambda-1}}
\end{equation}
Therefore the cumulant generating is (see also Fig.~\ref{Fig:U}) 
\begin{equation}
\label{U:sol}
U(\lambda) = - \log x_*(\lambda) = \log\frac{\sqrt{e^\lambda-1}}{\arctan\sqrt{e^\lambda-1}}
\end{equation}

Expanding $U(\lambda)$ in powers of $\lambda$, Eq.~\eqref{U-cum}, gives the leading behavior of the cumulants. We recover $\frac{\langle N\rangle}{L} = \frac{1}{3}$.  The ratios $F_k\equiv \langle\!\langle N^k\rangle\!\rangle/\langle N\rangle$ of cumulants to the average are known as Fano factors \cite{Fano}. The first four non-trivial Fano factors (by definition, $F_1=1$) are 
\begin{equation*}
F_2  = \frac{2}{15}\,, \quad 
F_3  = -\frac{2}{315}\,, \quad
F_4  =  -\frac{22}{1575}\,, \quad
F_5  = \frac{2}{1485}
\end{equation*}
The next five Fano factors are collected in Table~\ref{Table:6-10}.

\begin{table}[h!]
\centering
\renewcommand{\arraystretch}{1.5}{
\begin{tabular}{| c | c | c | c | c |}
\hline
$6$                                          & $7$                                      & $8$ & $9$  & $10$\\ 
\hline
$\frac{94\,442}{14\,189\,175}$  & $\frac{1\,622}{2\,027\,025}$  & $-\frac{3\,581\,702}{516\,891\,375}$ 
& $\frac{196\,599\,626}{206\,239\,658\,625}$  & $\frac{47\,221\,599\,182}{3\,781\,060\,408\,125}$ \\ 
\hline
\end{tabular}
}
\caption{Fano factors $F_k$ for $6\leq k\leq 10$.} 
\label{Table:6-10}
\end{table}

The Mandel $Q$ parameter \cite{Mandel79} defined via $Q=F_2-1$ is a basic measure characterizing the deviation from Poissonian statistics. The values  $-1\leq Q<\infty$ are permissible. For the Poisson statistics, $Q=0$; the sub-Poissonian range corresponds to $-1\leq Q<0$. Since $Q=-\frac{13}{15}$, the statistics of the final spatial arrangement of the houses in the RM is strongly sub-Poissonian. 

The Fano factors $F_k$ with $k\geq 2$ twice exceed the Fano factors appearing in the problem of random covering of the one-dimensional lattice by dimers \cite{Krapivsky23}. There is also a sign discrepancy: The periodic $(+,-,-,+)$ pattern for $F_k$ with $k\geq 2$ in the present case and the periodic $(+,+,-,-)$ pattern for $F_k$ with $k\geq 2$ in the covering by dimers \cite{Krapivsky23}. It would be interesting to understand the origin of this similarity.  

The method of computing the cumulants that we used in this subsection applies to other one-dimensional RSA processes \cite{Krapivsky20}. One deduces Riccati equations resembling \eqref{Phi:eq}, but apart from a few exceptional cases, these Riccati equations are unsolvable \cite{Krapivsky20}.

\subsection{Behavior at the boundary}
\label{subsec:boundary}

Let $e_L$ be the probability that in the jammed state, the leftmost site is empty. For small $L$, one can compute $e_L$ by hand:
$e_1=1,~ e_2=\frac{1}{2},~ e_3=\frac{1}{2}$, etc. Generally for $L\geq 2$, the quantities $e_L$ satisfy the recurrence
\begin{equation}
\label{eL:rec}
e_L =  \frac{1}{L}\sum_{k=1}^{L-1} e_k 
\end{equation}
Here $k+1$ is the site where the first house was built, with $k=1,\ldots,L-1$. The solution of the recurrence \eqref{eL:rec} exhibits complete saturation beginning from $L=2$:
\begin{equation}
\label{eL:sol}
e_1=1, \qquad e_L =  \frac{1}{2} \quad (L\geq 2)
\end{equation}

Let $e^{(2)}_L$ be the probability that in the jammed state, the 2nd site is empty. One finds 
$e^{(2)}_2=\frac{1}{2},~ e^{(2)}_3=\frac{1}{3}$, etc. Generally for $L\geq 2$, the quantities $e^{(2)}_L$ satisfy the recurrence
\begin{equation}
\label{e2L:rec}
e^{(2)}_L =  \frac{e_{L-1}}{L}+ \frac{1}{L}\sum_{k=2}^{L-1} e^{(2)}_k 
\end{equation}
The solution of the recurrence \eqref{e2L:rec} exhibits complete saturation beginning from $L=3$:
\begin{equation}
\label{e2L:sol}
e^{(2)}_2=\frac{1}{2}, \qquad e^{(2)}_L =  \frac{1}{3} \quad (L\geq 3)
\end{equation}

Similarly the probabilities $e^{(3)}_L$ that in the jammed state the 3rd site is empty satisfy
\begin{equation}
\label{e3L:rec}
e^{(3)}_L =  \frac{e^{(2)}_{L-1}}{L}+ \frac{e_{L-2}}{L}+ \frac{1}{L}\sum_{k=3}^{L-1} e^{(3)}_k 
\end{equation}
The solution of the recurrence \eqref{e3L:rec} exhibits complete saturation beginning from $L=3$:
\begin{equation}
\label{e3L:sol}
e^{(3)}_3=\frac{1}{2}, \qquad e^{(3)}_L =  \frac{1}{3} \quad (L\geq 4)
\end{equation}

The complete saturation can be extended to sites further away from the boundary. Overall, the jamming density $\rho_j$ at the site $j$ is
\begin{equation}
\label{jam:segment}
\rho_1=\rho_L=\frac{1}{2}, \qquad \rho_j=\frac{2}{3} \quad (1<j<L)
\end{equation}
The same saturation phenomenon occurs on the semi-infinite one-dimensional lattice, see \eqref{jam:semi}.

\subsection{Extremal jammed configurations}
\label{subsec:min-max}

Let $M_L$ be the probability of creating the maximally dense jammed configuration, i.e., a configuration of the type \eqref{dense}. This happens if houses are consecutively built on the boundaries. Therefore
\begin{equation}
\label{ML-exact}
M_L = \prod_{j=2}^{L} \frac{2}{j} = \frac{2^{L-1}}{L!}
\end{equation}

Suppose $L$ is odd. In this case, there is a unique maximally sparse jammed configuration \eqref{sparse:odd}. We write $L=2n+1$ and denote by $m_n$ the probability of creating the jammed configuration \eqref{sparse:odd}. The probability of reaching this configuration can be determined from recurrence 
\begin{equation}
\label{m-n}
m_n = \frac{1}{2n+1}\sum_{k=0}^{n-1}m_k m_{n-k-1}
\end{equation}
The boundary condition is $m_0 =  1$. The same recurrence appeared in \cite{Krapivsky23} in the case of covering by dimers. The generating function 
\begin{equation}
\label{m:GF}
\sum_{k\geq 0}m_k x^{k} = \frac{\tan \sqrt{x}}{\sqrt{x}}
\end{equation}
was computed in \cite{Krapivsky23}. The generating function \eqref{m:GF} has a simple pole at $x=(\pi/2)^2$. This allows one to extract the large $n$ asymptotic
\begin{equation}
\label{mu-asymp}
m_n \simeq 2\left(\frac{2}{\pi}\right)^{2n+2}  = \frac{4}{\pi}\left(\frac{2}{\pi}\right)^L
\end{equation}

\section{Riviera Model with egoistic settlers on the line}
\label{sec:line}

In this section, we analyze the continuum version of the RM defined as follows:
\begin{enumerate}
\item The coastline is a line. 
\item The houses are identical, each having a facade of length $w$. 
\item A new house can be built in an empty interval if the separation from at least one of the adjacent houses exceeds a threshold distance $\ell$. 
\item Each new house is built instantaneously. 
\end{enumerate}
The 3rd condition ensures that a new house initially enjoys sufficient sunlight, although it may disappear as the settlement continues. 

When $\ell=0$, we recover a classical car-parking problem introduced by R\'{e}nyi \cite{Renyi58}. The case $\ell>0$ leads to substantially more involved calculations, although the methods developed for the car-parking problem extend to it. The ratio $\ell/w$ is the relevant dimensionless parameter. For simplicity we consider the process with ratio equal to one. Without loss of generality we can set $w=1$, i.e., measure the length in the unit of facade length. Then $w=\ell=1$ when the ratio is equal to one. 

Denote by $V(x,t)$ the density of voids of length $x$, i.e., the probability density of the patterns
\begin{equation}
\label{void:line}
\Box \underbrace{\qquad\qquad}_{x}\Box
\end{equation}
where $\Box$ denotes a house. The density $V(x,t)$ satisfies 
\begin{subequations}
\begin{align}
\label{long}
\frac{\partial V(x,t)}{\partial t} = -(x-1)V(x,t)+2\int_{x+1}^\infty dy\,V(y,t)
\end{align}
when the void is long, $x>3$. The first term on the RHS of Eq.~\eqref{long} accounts for building a house inside the void. The second term on the RHS of Eq.~\eqref{long} describes the creation of an $x$-void by building a house in voids of length $\geq x+1$; there are exactly two proper locations. Similarly we have 
\begin{align}
\label{medium}
\frac{\partial V(x,t)}{\partial t} = -2(x-2)V(x,t)+ 2\int_{x+1}^\infty dy\,V(y,t)
\end{align}
for voids of intermediate length, $3>x>2$. Shorter voids can only be created. Their densities obey 
\begin{align}
\label{12}
&\frac{\partial V(x,t)}{\partial t} =  2\int_{x+1}^\infty dy\,V(y,t), \quad  1<x<2\\
\label{01}
&\frac{\partial V(x,t)}{\partial t} =  2\int_{x+2}^\infty dy\,V(y,t), \quad  x<1
\end{align}
\end{subequations}

To solve \eqref{long} we employ the ans\"{a}tze 
\begin{equation}
\label{V:ansatz}
V(x,t) = e^{-(x-1)t}\,\Phi(t), \qquad x>3
\end{equation}
and reduce \eqref{long} to 
\begin{equation}
\label{Phi-eq}
\frac{d \Phi(t)}{d t} = \frac{2 e^{-t}\Phi(t)}{t}
\end{equation}
Integrating \eqref{Phi-eq} subject to $\Phi(0)=0$ we obtain 
\begin{equation}
\label{Phi-sol}
\Phi(t) =  t^2 \exp\!\left[-2\int_0^t d\tau\,\frac{1-e^{-\tau}}{\tau}\right]
\end{equation}
The asymptotic behaviors are
\begin{equation}
\label{Phi-asymp}
\Phi(t) \simeq
\begin{cases}
  t^2                    & t\to 0\\
  e^{-2\gamma}   & t\to\infty
\end{cases}  
\end{equation}
where $\gamma=0.57721566\ldots$ is the Euler constant. The large time behavior becomes obvious after re-writing \eqref{Phi-sol} as 
\begin{equation}
\label{Phi-gamma}
\Phi(t)  =  \exp\!\left[-2\gamma-2\Gamma(0,t)\right]
\end{equation}
where $\Gamma(0,t)=\int_t^\infty dy y^{-1}e^{-y}$ is the incomplete gamma function.

Using \eqref{V:ansatz} we compute the integral in \eqref{medium} and obtain 
\begin{align}
\label{medium:V}
\left[\frac{\partial}{\partial t} + 2(x-2)\right]V(x,t) = \frac{2\Phi(t)}{t}\,e^{-xt}
\end{align}
which is integrated to yield
\begin{equation}
\label{Vxt:medium}
V(x,t)  =  2e^{-xt}\int_0^t d\tau\,\frac{\Phi(\tau)}{\tau}\,e^{(4-x)(t-\tau)}
\end{equation}
when $2<x<3$. Similarly, we reduce Eq.~\eqref{12} to
\begin{eqnarray}
\label{V12}
\frac{\partial V(x,t)}{\partial t} &=&  4e^{-2t}\int_0^t d\tau\,\frac{e^{-\tau}\Phi(\tau)}{\tau}\,\frac{e^{(2-x)(2t-\tau)}-1}{2t-\tau}  \nonumber \\
&+& \frac{2 e^{-2t}\Phi(t)}{t}
\end{eqnarray}
and Eq.~\eqref{01} to
\begin{eqnarray}
\label{V01}
\frac{\partial V(x,t)}{\partial t} &=&  4e^{-2t}\int_0^t d\tau\,\frac{e^{-\tau}\Phi(\tau)}{\tau}\,\frac{e^{(1-x)(2t-\tau)}-1}{2t-\tau} \nonumber \\
&+& \frac{2 e^{-2t}\Phi(t)}{t} 
\end{eqnarray}
The density of voids of length $1<x<2$  is found by integrating \eqref{V12}  subject to $V(x,0)=0$. Similarly, the density of voids of length $0<x<1$ is found by integrating \eqref{V01} subject to $V(x,0)=0$.

The fraction of empty space is $\int_0^\infty dx\,x V(x,t)$. Using \eqref{V:ansatz} we compute the contribution from long voids 
\begin{equation}
\int_3^\infty dx\,x V(x,t) =  \frac{(1+3t) e^{-2t}\Phi(t)}{t^2}
\end{equation}
Using Eq.~\eqref{Phi-asymp} we find that this contribution decays as $3 e^{-2\gamma} t^{-1} e^{-2t}$ when $t\gg 1$. 

The contribution from medium voids 
\begin{eqnarray}
\int_2^3 dx\,x V(x,t) &=&  2\int_0^t d\tau\,\frac{\Phi(\tau)}{\tau}\,\frac{2e^{-2\tau}-3e^{-2t-\tau}}{2t-\tau} \nonumber\\
&+ & 2\int_0^t d\tau\,\frac{\Phi(\tau)}{\tau}\,\frac{e^{-2\tau}-e^{-2t-\tau}}{(2t-\tau)^2}
\end{eqnarray}
decays as $Ct^{-1}$, where $C=2\int_0^\infty d\tau\,\frac{e^{-2\tau}\Phi(\tau)}{\tau}$. The numerical value is $C\approx0.149\,939$. 

The voids of length $0<x<1$ and $1<x<2$ have the fraction of empty space 
\begin{equation}
e(t)=\int_0^1 dx\,x V(x,t) + \int_1^2 dx\,x V(x,t)
\end{equation}
Using \eqref{V12} and \eqref{V01} we find that this fraction increases with time according to 
\begin{eqnarray}
\frac{de}{dt} = 4\frac{e^{-2t} \Phi(t)}{t} + \int_0^t d\tau\,\frac{e^{-2\tau}\Phi(\tau)}{\tau}\,F(2t-\tau)
\end{eqnarray}
where  
\begin{equation}
F(T)=\frac{T+2-(2+3T+2T^2)e^{-T}}{T^3}
\end{equation}
The fraction $e_\infty=e(t=\infty)$ of empty space in the jammed state is
\begin{eqnarray}
e_\infty =2 \int_0^\infty dt\,\frac{e^{-2t}\Phi(t)}{t^3}[1+t+2t^2-(1+2t)e^{-t}]
\end{eqnarray}
The numerical value is $e_\infty \approx 0.468\,583$. 

The continuum RM with $\ell=0$ reduces to the car-parking problem. The fraction of empty space is \cite{Evans93,Talbot00,KRB} 
\begin{equation}
e(t) = 1 - \int_0^t dv\,\exp\!\left[-2\int_0^v d\tau\,\frac{1-e^{-\tau}}{\tau}\right]
\end{equation}
In the jammed state \cite{Renyi58}
\begin{equation}
e_\infty = 1 - \int_0^\infty dv\,\exp\!\left[-2\int_0^v d\tau\,\frac{1-e^{-\tau}}{\tau}\right]
\end{equation}
The numerical value is $e_\infty\approx 0.252\,402$.

\section{Discussion}
\label{sec:disc}

We obtained an exact characterization of the RM with egoistic settlers on the infinite one-dimensional lattice. One of our central results is the cluster-size distribution in the jammed state, Eq.~\eqref{Pi-s}. Although the cluster densities do not form a closed hierarchy, the void-cluster-void distribution does, and an exponential ans\"{a}tze reduces its evolution to a recursive hierarchy of linear ODEs. The derivation of Eq.~\eqref{Pi-s} is thus indirect and required us to derive much more comprhensive time-dependent predictions for the void-cluster-void distribution. 

The direct probe of some characteristics of the jammed states in finite systems is still feasible (Sec.~\ref{sec:segment}). For instance, we computed the statistics of the number of empty sites $N$ in the jammed states on segments of length $L$. In Sec.~\ref{subsec:FCS}, we computed the large $L$ behavior of the all cumulants $\langle\!\langle N^n\rangle\!\rangle$. In Sec.~\ref{subsec:boundary}, we demonstrated a remarkable saturation property of the jamming density: The jamming density is the same at all sites apart from the boundary sites. We also explored the continuum version of the RM, Sec.~\ref{sec:line}. 

Throughout this article, we investigated the evolution of initially empty system. If the system is partially occupied but the initial state is compatible with an exponential ans\"{a}tze, analytical progress is feasible. The void distributions in several one-dimensional RSA processes in such initial states have been computed \cite{BK94}, and it should be possible to perform such computations for the RM. 

Extensions to more complicated distributions appear feasible. Suppose we want to compute the cluster-cluster-size distribution describing the densities of adjacent clusters of prescribed sizes. One must also specify the size of the void between the clusters, so the relevant cluster-void-cluster patterns are  
\begin{equation}
\circ\, \underbrace{\bullet \cdots \bullet}_{s_1} \underbrace{\circ \cdots \circ}_{n} \underbrace{\bullet \cdots \bullet}_{s_2}\circ 
\end{equation}
The corresponding densities $P_{s_1,s_2}(n; t)$ do not satisfy a closed set of equations. To overcome this problem, one should study the void-cluster-void-cluster-void distribution. The densities $P_{s_1,s_2}(m_1,m_2, m_3; t)$ of patterns
\begin{equation}
\bullet \underbrace{\circ \cdots \circ}_{m_1}  \underbrace{\bullet \cdots \bullet}_{s_1} \underbrace{\circ \cdots \circ}_{m_2} 
\underbrace{\bullet \cdots \bullet}_{s_2}  \underbrace{\circ \cdots \circ}_{m_3} \bullet 
\end{equation}
satisfy a closed set of equations that can presumably be solved with the help of the exponential ans\"{a}tze. In the jammed state, only patterns with minimal voids are present:
\begin{equation}
\bullet \circ \underbrace{\bullet \cdots \bullet}_{s_1} \circ \underbrace{\bullet \cdots \bullet}_{s_2}  \circ\,\bullet 
\end{equation}
i.e., $\Pi_{s_1,s_2}\equiv P_{s_1,s_2}(1,1, 1; \infty)$. Computing $\Pi_{s_1,s_2}$ and more complicated densities $\Pi_{s_1,s_2,s_3}$ of patterns
\begin{equation}
\bullet \circ \underbrace{\bullet \cdots \bullet}_{s_1} \circ \underbrace{\bullet \cdots \bullet}_{s_2}  \circ  \underbrace{\bullet \cdots \bullet}_{s_3} \circ\,\bullet 
\end{equation}
should be possible. Such calculations appear straightforward but are exceedingly laborious due to the necessity of considering joint distributions $P_{\bf s}({\bf m};t)$, with $k$ clusters ${\bf s}=(s_1,\ldots,s_k)$ of arbitrary sizes, $s_i\geq 1$, and $(k+1)$ voids ${\bf m}=(m_1,\ldots,m_{k+1})$  of arbitrary sizes, $m_j\geq 1$. In the jammed state, only the densities $\Pi_{\bf s}\equiv P_{\bf s}({\bf 1};\infty)$ with ${\bf m}={\bf 1}=(1,\ldots,1)$ are non-vanishing.

We also note an intriguing relation between the Fano factors associated with the number of empty sites in the jammed state and the Fano factors describing the statistics of the number of dimers in the dimer random covering process \cite{Krapivsky23}. The discrepancies (an extra factor of two and the sign pattern) may disappear if, instead of empty sites, one considers occupied sites. 

Random covering processes with polymers longer than dimers are much more complicated \cite{Krapivsky23}. For infinite systems, explicit exact solutions for trimers, tetramers, and pentamers have been found \cite{Pascal25};  generalization to longer polymers is feasible but very cumbersome \cite{Pascal25}. For finite systems, only random coverings by dimers yielded to analytical treatment \cite{Krapivsky23}. Thus, the hidden isomorphism between the RM and random coverings by dimers hints at the possible existence of processes resembling the RM that are isomorphic to random coverings by trimers, tetramers, pentamers, and longer polymers.

\bigskip\noindent
I am grateful to Jean-Marc Luck for useful remarks. 

\appendix

\section{Riviera model with egoistic settlers on semi-infinite one-dimensional lattice}
\label{sec:one-sided}

Here we briefly consider the RM process on the semi-infinite lattice $\mathbb{Z}_+=\{j\in \mathbb{Z}, j\geq 1\}$. Suppose the process begins with an initially empty semi-infinite lattice
\begin{equation}
\label{IC:semi}
\blacktriangleleft\circ\circ\ldots
\end{equation}
where we use $\blacktriangleleft$ to indicate that the leftmost site does not receive sunlight from the outside. The spatial inhomogeneity makes the process on   $\mathbb{Z}_+$ significantly more challenging than the RM process on $\mathbb{Z}$. In the RM process on the infinite one-dimensional lattice, it was easier to deal with the densities $W_n$ of $n$ consecutive empty sites than with the densities $V_n$ of voids of length $n$. The same is true for the semi-infinite lattice. Because we are mostly interested in the evolution near the boundary, the simplest analogs of $W_n$ are the densities
\begin{equation}
\label{Bn:def}
B_n(t) = \text{Prob}\big[\!\blacktriangleleft\underbrace{\circ \cdots \circ}_{n}\big]
\end{equation}
Equivalently, $B_n(t)$ is the probability that $n$ consecutive sites at the boundary of $\mathbb{Z}_+$ are empty at time $t$. The densities $B_n(t)$ evolve according to equations
\begin{subequations}
\label{B:voids+}
\begin{align}
\label{B:void-n+}
\frac{dB_n}{dt} &=-n B_n, \qquad n\geq 2\\
\label{B:void-1+}
\frac{dB_1}{dt} &=-B_2
\end{align}
\end{subequations}
derived similarly to Eqs.~\eqref{voids+}. Indeed, the evolution of $B_n$ with $n\geq 2$ is independent of the state of the $(n+1)^{\text{st}}$ site explaining \eqref{B:void-n+}. The evolution of $B_1$ is possible, in the case when the initial condition is given by \eqref{IC:semi}, only if the 2nd site is empty leading to \eqref{B:void-1+}. 

Solving \eqref{B:void-n+} and then \eqref{B:void-1+} subject to $B_n(0)=1$ for all $n\geq 1$ we obtain
\begin{subequations}
\label{Bn:sol}
\begin{align}
\label{Bn+:sol}
&B_n = e^{-nt}\,, \qquad n\geq 2\\
\label{B1+:sol}
&B_1 = \frac{1+e^{-2t}}{2}
\end{align}
\end{subequations}
Thus $\rho_1(t)=1-B_1=\frac{1-e^{-2t}}{2}$ is the probability that the first site is occupied. The jamming density $\rho_1(\infty)=\frac{1}{2}$
is smaller than the jamming density $\rho_\text{jam}=\frac{2}{3}$ in the bulk. 

To probe the fate of the second site, we employ the densities
\begin{equation}
\label{Bn1:def}
B^{(1)}_n(t) = \text{Prob}\big[\!\blacktriangleleft\star\underbrace{\circ \cdots \circ}_{n}\big]
\end{equation}
with $\star$ implying that the state (empty or occupied) of the leftmost site is unspecified.  These densities satisfy
\begin{subequations}
\label{B1:voids+}
\begin{align}
\label{B1:void-n+}
\frac{dB^{(1)}_n}{dt} &=-n B^{(1)}_n, \qquad n\geq 2\\
\label{B1:void-1+}
\frac{dB^{(1)}_1}{dt} &=-B_2-\big(B^{(1)}_2-B_3\big)
\end{align}
\end{subequations}
The evolution equations \eqref{B1:void-n+} are obvious. The density $B^{(1)}_1$ evolves when the first two sites are in the state $\circ \circ$, i.e., with rate $B_2$, and if the first three sites are in the state $\bullet \circ \circ$, i.e., with rate $B^{(1)}_2-B_3$. 

Solving \eqref{B1:void-n+} and then \eqref{B1:void-1+} subject to $B^{(1)}_n(0)=1$ for all $n\geq 1$ we obtain
\begin{subequations}
\label{B1n:sol}
\begin{align}
\label{B1n+:sol}
&B^{(1)}_n = e^{-nt}\,, \qquad n\geq 2\\
\label{B1+:sol}
&B^{(1)}_1 = \frac{1-e^{-3t}}{3}+e^{-2t}
\end{align}
\end{subequations}
We conclude that the second site is occupied with probability $\rho_2(t)=1-B^{(1)}_1=\frac{2+e^{-3t}}{3}-e^{-2t}$. The jamming density of the second site 
$\rho_2(\infty)=\frac{2}{3}$ is the same as the jamming density in the bulk. 

We now consider the densities
\begin{equation}
\label{Bnk:def}
B^{(k)}_n(t) = \text{Prob}\big[\!\blacktriangleleft \underbrace{\star\  \cdots \star}_k \, \underbrace{\circ \cdots \circ}_{n}\big]
\end{equation}
with arbitrary $k\geq 2$. These densities satisfy
\begin{subequations}
\label{Bk:voids+}
\begin{align}
\label{Bk:void-n+}
\frac{dB^{(k)}_n}{dt} &=-n B^{(k)}_n, \qquad n\geq 2\\
\label{Bk:void-1+}
\frac{dB^{(k)}_1}{dt} &=-B^{(k-1)}_2-\big(B^{(k)}_2-B^{(k-1)}_3\big)
\end{align}
\end{subequations}
which are solved to yield
\begin{subequations}
\label{Bkn:sol}
\begin{align}
\label{Bkn+:sol}
&B^{(k)}_n = e^{-nt}\,, \qquad n\geq 2\\
\label{Bk+:sol}
&B^{(k)}_1 = \frac{1-e^{-3t}}{3}+e^{-2t}
\end{align}
\end{subequations}
identical to \eqref{B1n:sol}. This saturation property, i.e., the lack of dependence on $k$ for all $k\geq 1$, implies the saturation of the jamming density to the bulk value for all sites apart from the leftmost site:
\begin{equation}
\label{jam:semi}
\rho_1(\infty)=\frac{1}{2}\,, \qquad \rho_j(\infty)=\frac{2}{3} \quad (j\geq 2)
\end{equation}

\section{Generalized Riviera model with egoistic settlers}
\label{ap:gen}

The original interpretation of the Riviera model suggests that an empty site with two empty neighbors should be more favorable for building than an empty site with only one empty neighbor. This observation suggests extending the RM to a one-parameter class of models defined as follows:
\begin{itemize}
\item Each site is chosen independently with a rate that we again set to one.
\item The house is built if the site is empty and both neighboring sites are empty. 
\item The house is built with probability $\epsilon$ if the site and one neighboring site are empty. 
\item An attempt is rejected if there is already a house in the site, or the site is empty but surrounded by two occupied sites. 
\end{itemize}

Some results for the RM with $\epsilon=1$, Secs.~\ref{sec:evol}--\ref{sec:segment}, can be generalized to the RM with arbitrary $\epsilon$. (The models with $\epsilon>1$ are also admissible after one interpretes $\epsilon$ as a rate rather than a probability.) 

The void distribution can be computed explicitly for arbitrary $\epsilon$. Instead of Eqs.~\eqref{voids}--\eqref{void-1} we have 
\begin{subequations}
\label{VV:eq}
\begin{align}
\label{Vn:eq}
\frac{dV_n}{dt} &=-(n-2+2\epsilon) V_n +2\epsilon V_{n+1}+2\sum_{j\geq n+2}V_j\\
\label{V1:eq}
\frac{dV_1}{dt} &=2\epsilon V_2 + 2\sum_{j\geq 3}V_j
\end{align}
\end{subequations}

Using the exponential ans\"{a}tze \eqref{exp} we reduce the infinite system \eqref{Vn:eq} to ODEs for $a(t)$ and $A(t)$. The former quantity satisfies the same Eq.~\eqref{a:eq} as before, and therefore the solution is again
$a = e^{-t}$. The amplitude obeys 
\begin{equation}
\label{Aa-gen}
A^{-1}\,\frac{dA}{da} = 2(1-\epsilon)\left[1-\frac{1}{a}\right]-\frac{2}{1-a}
\end{equation}
When $\epsilon=1$, we recover Eq.~\eqref{Aa:eq}.

Integrating \eqref{Aa-gen} subject to \eqref{A:IC} we find 
\begin{equation}
\label{A-exp}
A=\frac{(1-a)^2}{a^{2(1-\epsilon)}}\,\mathcal{E}(a), \quad  \mathcal{E}(a)=e^{-2(1-\epsilon)(1-a)}
\end{equation}

We now substitute the exponential ans\"{a}tze \eqref{exp} with $A$ given by \eqref{A-exp} into \eqref{V1:eq} and find
\begin{equation}
\label{V1a:eq}
\frac{dV_1}{da} = -2(1-a)a^{2\epsilon-1}[\epsilon(1-a)+a]\mathcal{E}(a)
\end{equation}
Integrating \eqref{V1a:eq} subject to the initial condition reflecting that we start with an empty system, $V_1(a=1)=0$, gives an exact solution. We do not write the cumbersome expression of $V_1(a)$ and cite only the final jammed value, the fraction $e_\infty(\epsilon)\equiv V_1(t=\infty; \epsilon)$ of empty sites in the jammed state:
\begin{subequations}
\label{E-epsilon}
\begin{equation}
e_\infty(\epsilon) = 2^{-2 \epsilon -1} e^{2 \epsilon -2} (\epsilon -1)^{-2 \epsilon -1} W(\epsilon)
\end{equation}
where
\begin{eqnarray}
W(\epsilon) &=& (4 \epsilon -2)  \Gamma[2 \epsilon +1, 2\epsilon -2]-\Gamma[2 \epsilon +2, 2\epsilon -2]\nonumber \\
&+&    4\epsilon  (1-\epsilon ) \Gamma[2 \epsilon, 2\epsilon -2] + \Gamma (2 \epsilon +1)
\end{eqnarray}
\end{subequations}

\begin{figure}[ht]
\begin{center}
\includegraphics[width=0.44\textwidth]{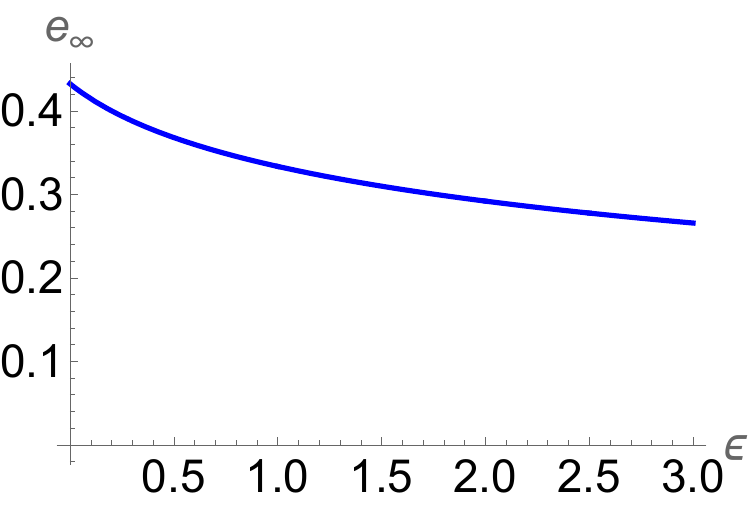}
\caption{The plot of the density of empty sites $e_\infty$ versus $\epsilon$ in the jammed state. The analytical expression for $e_\infty(\epsilon)$ is given by Eqs.~\eqref{E-epsilon}. At the minimal, intermediate, and maximum values of the probability $\epsilon$, the jammed density is  $e_\infty(+0)=\frac{1-e^{-2}}{2}$, $e_\infty(\frac{1}{2})=e^{-1}$, and $e_\infty(1)=\frac{1}{3}$. When $\epsilon>1$, we interpret $\epsilon$ as a rate. }
\label{Fig:Empty}
  \end{center}
\end{figure}

The jammed density of the empty sites $e_\infty(\epsilon)$ is a decreasing function of $\epsilon$, see Fig.~\ref{Fig:Empty}. The largest density $e_\infty(+0)=\frac{1-e^{-2}}{2}$ follows from Eqs.~\eqref{E-epsilon}, and it can be alternatively deduced from a connection with the simplest RSA model known as unfriendly seating arrangement process \cite{Shepp}. In this RSA model, each occupied seat is surrounded by two empty seats, so the final jammed configuration locally looks like  
\begin{equation}
\label{unfriendly}
\cdots \circ\,\bullet\,\circ\,\bullet\,\circ\,\circ\,\bullet\,\circ\,\bullet\,\circ\,\bullet\,\circ\,\circ\,\bullet\,\circ\,\bullet\,\circ\,\circ\,\bullet\,\circ\cdots
\end{equation}

In the unfriendly seating arrangement process, the jammed density of occupied seats is $\frac{1-e^{-2}}{2}$, so the density of empty seats is $\frac{1+e^{-2}}{2}$. In the RM with $\epsilon=+0$, the system first reaches a configuration like \eqref{unfriendly} corresponding to $\epsilon=0$, and then voids of length two evolve independently and become voids of length one. The jammed density equals the density of empty seats in the model of unfriendly seating arrangement, i.e., $\frac{1+e^{-2}}{2}$; the density of empty sites is $\frac{1-e^{-2}}{2}$. There is also a relation with the classical RSA model \cite{Flory39} describing the adsorption of dimers  in which the density of occupied sites is $1-e^{-2}$. In addition to $\epsilon=1$ when $e_\infty(1)=\frac{1}{3}$, the choice $\epsilon=\frac{1}{2}$ is also natural: The house is built in an empty site with rate equal to the fraction of adjacent empty sites. In this case, Eqs.~\eqref{E-epsilon} yield $e_\infty(\frac{1}{2})=e^{-1}$. 

The models with $\epsilon>1$ make sense if one interprets $\epsilon$ as a rate rather than a probability. If $\epsilon>1$ is half-integer, the jammed density has a neat form: 
\begin{equation*}
e_\infty\!\left(\frac{k}{2}+1\right)=\frac{(k+2)! e^k - 8A_k}{k^{k+3}}
\end{equation*}
with $A_k=2, 21, 276, 4370, 81030, 1722672, 41311592$ for $k=1,\ldots,7$. (Using \eqref{E-epsilon}, one can express $A_k$ via incomplete gamma functions. We haven't succeeded in proving that $A_k$ are all integers.)

The same arguments as in Sec.~\ref{sec:PCF} give the connected pair correlation functions $C_j(\epsilon)$ with $|j|\leq 3$
\begin{equation}
\label{C0123}
\begin{split}
& C_0(\epsilon) = e_\infty(\epsilon)[1-e_\infty(\epsilon)]\\
& C_1(\epsilon) = -[e_\infty(\epsilon)]^2  \\
& C_2(\epsilon) = \Pi_1(\epsilon) -[e_\infty(\epsilon)]^2\\
& C_3(\epsilon) = \Pi_2(\epsilon) -[e_\infty(\epsilon)]^2
\end{split}
\end{equation}
 in the jammed state. Equations \eqref{C0123} are valid for arbitrary $\epsilon$. The cluster densities $\Pi_1(\epsilon)$ and $\Pi_2(\epsilon)$ are known when $\epsilon=1$. Even in this special case, $\Pi_1$ and $\Pi_2$ do not satisfy a closed set of equations --- we needed to determine the entire distribution $\Pi_s$. (The distribution $\Pi_s$ was actually extracted from the more comprehensive void-cluster-void distribution.) Thus when $\epsilon>0$ and $\epsilon\ne 1$, we only know two connected pair correlation functions: $C_0(\epsilon)$ and $C_1(\epsilon)$. 
 
 In addition to $\epsilon=1$, there is another exceptional case: $\epsilon=+0$. In this case, $\Pi_s(+0)=0$ for $s\geq 3$, and hence
\begin{equation*}
\begin{split}
&1=2\Pi_1(+0)+3\Pi_2(+0)\\
&1-e_\infty(+0)=\Pi_1(+0)+2\Pi_2(+0)\
\end{split}
\end{equation*}
Recalling $e_\infty(+0)=\frac{1-e^{-2}}{2}$ we fix 
\begin{equation}
\Pi_1(+0)=\frac{1-4e^{-2}}{2}\,, \qquad \Pi_2(+0) = e^{-2}
\end{equation}
Therefore, Eqs.~\eqref{C0123} give
\begin{equation}
\label{C0123:0}
\begin{split}
& C_0(+0) = \frac{1-e^{-4}}{4}\\
& C_1(+0) = -\left(\frac{1-e^{-2}}{2}\right)^2  \\
& C_2(+0) = \frac{1-4e^{-2}-e^{-4}}{4}\\
&  C_3(+0) = \frac{4e^{-2}-1-e^{-4}}{2} 
\end{split}
\end{equation}
Thus, the connected pair correlation functions $C_j(\epsilon)$ with $|j|\leq 3$ is explicitly known when $\epsilon=+0$ and $\epsilon=1$. In the latter case, the available exact results are compatible with complete decorrelation, $C_j(1)=0$ for $|j|\geq 3$. For the model with $\epsilon=+0$, we have $C_3(+0)\approx -0.238487$. 

For the model with $\epsilon=1$, we established complete saturation of the jamming density one site away from the boundaries, cf. \eqref{jam:segment} and \eqref{jam:semi}. This saturation property is reminiscent of complete decorrelation. Complete decorrelation is apparently lacking in all models with $\epsilon\ne 1$. Complete saturation of the jamming density may also be violated in models with $\epsilon\ne 1$, and it could be possible to check in the rather tractable model with $\epsilon=+0$. Computing the variance of the number of empty sites in the model with $\epsilon=+0$ is another interesting challenge. This would give us the sum $\sum_{-\infty<j<\infty}C_j(+0)$, see \eqref{Cj:sum-var}. The variance is known \cite{Krapivsky20} for the model with $\epsilon=0$, which describes the unfriendly seating arrangement process. The connection between the models with $\epsilon=0$ and $\epsilon=+0$ hints that the variances in the two models could be related. If such a relation exists, it is not as straightforward as the relation between the jamming densities that we used above. 

\section{Configurational entropy}
\label{ap:entropy}

Here, we disregard dynamics and consider jammed configurations on the segment of length $L$. The total number of jammed configurations $J_L$ satisfies the recurrence $J_L = J_{L-1}+J_{L-2}$ which is readily understood after dividing the jammed states into the complementary sets with the occupied and empty leftmost site. Therefore $J_L$ is a Fibonacci number; $J_L=F_{L+1}$ if we use the standard definition of the Fibonacci sequence. The growth of  $J_L$ with $L$ is exponential: $J_L\sim g^L$, where $g=\frac{1+\sqrt{5}}{2}$ is the golden ratio. 

The total number $J(L,H)$ of jammed configurations with a fixed number of houses $H$ also exhibits an exponential growth:
\begin{equation}
\label{sigma-def}
\lim_{L\to\infty} \frac{\log J(L,H)}{L}=\sigma(\rho)
\end{equation}
in the thermodynamic limit
\begin{equation}
\label{TL}
L\to\infty, \quad H\to\infty, \quad \rho=\frac{H}{L}=\text{fixed}
\end{equation}

\begin{figure}[ht]
\begin{center}
\includegraphics[width=0.44\textwidth]{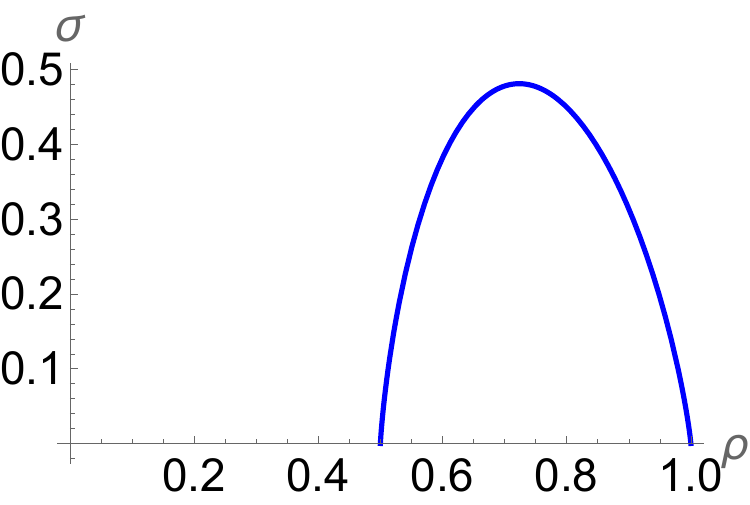}
\caption{The configurational entropy $\sigma$ versus the density $\rho$ of houses in the jammed state. The analytical expression for $\sigma(\rho)$ is given by Eq.~\eqref{sigma}. The entropy reaches maximum 
$\sigma_*=\sigma(\rho_*) = \text{ArcCoth}\big[\sqrt{5}\big]\approx 0.481212$  
at the density $\rho_*=\frac{5+\sqrt{5}}{10}\approx 0.723607$. }
\label{Fig:entropy}
  \end{center}
\end{figure}

The configurational entropy $\sigma(\rho)$ is 
\begin{equation}
\label{sigma}
\sigma(\rho) = \rho \log \rho - (1-\rho) \log1-\rho)- (2\rho-1)\log(2\rho-1)
\end{equation}
in the present case. The maximum (see Fig.~\ref{Fig:entropy}) is reached at $\rho_*=\frac{5+\sqrt{5}}{10}\approx 0.723607$. Comparing this density with the jamming densities $\rho_\text{jam}(\epsilon)=1-e_\infty(\epsilon)$ for the class of models analyzed in Appendix \ref{ap:gen} we conclude that $\rho_\text{jam}(\epsilon)<\rho_*$ when $\epsilon < \epsilon_*$ and $\rho_\text{jam}(\epsilon)>\rho_*$ when $\epsilon > \epsilon_*$. The threshold value, $\epsilon_*\approx 2.541181761$, is the root of 
\begin{equation}
e_\infty(\epsilon_*)=1-\rho_* = \frac{5-\sqrt{5}}{10}
\end{equation}

The determination of $J(L,H)$ is a neat problem with combinatorial flavor---one seeks configurations with all empty sites isolated, and their number fixed and equal to $L-H$. This combinatorial problem has arisen in many studies, e.g., in Refs.~\cite{Alan10,Krap13,Moore}. The exact answer 
\begin{equation}
\label{J-LH}
J(L,H) = \binom{H}{L-H}+\binom{H-1}{L-H-1}
\end{equation}
holds when $H\geq L/2$. In the exceptional case of $H=k$ and $L=2k+1$, we have $J(2k+1,k) = 1$ representing the maximally sparse jammed configuration \eqref{sparse:odd}. The number of maximally dense jammed configurations predicted by  \eqref{J-LH} is $J(L,L-1)=L$, in agreement with the previous direct calculation relying on \eqref{dense}. Equation \eqref{J-LH} leads to \eqref{sigma-def} in the thermodynamic limit \eqref{TL}, with configurational entropy $\sigma(\rho)$ given by \eqref{sigma}. 

The computation of the configurational entropy, also known as complexity \cite{Moore}, is simpler than solving the dynamical process. For instance, the dynamical version of the original Riviera model has not been solved, while the configurational entropy has been computed \cite{Puljiz24,JM23a}. In the original Riviera model, the allowed range of density is $\frac{1}{2}<\rho<\frac{2}{3}$; the entropy reaches maximum at the density that is analytically known, $\rho_* \approx 0.577203$. The jamming density $\rho_\text{jam}\approx 0.600385$ is known only numerically \cite{JM23a}. 

One can deform the original Riviera model in a manner similar to the deformation of the RM described in Appendix~\ref{ap:gen}. Namely, one postulates that $\circ\circ\circ \Longrightarrow \circ\bullet\circ$ proceeds with rate set to one, while the rate of the process $\circ\circ\bullet\,\circ \Longrightarrow \circ\bullet\bullet\,\circ$ is $\epsilon$; other processes are forbidden as the maximal cluster size is two. The version studied so far \cite{Puljiz24,JM23a} corresponds to $\epsilon=1$. The version with $\epsilon=+0$ could be amenable to analytical progress. One can appreciate the upper bound for jamming density, $\rho_\text{jam}(+0)<\frac{1+e^{-2}}{2}\approx 0.567668$, by realizaing that the system first reaches a quasi-jammed state like \eqref{unfriendly}, and then some of the voids of length two become voids of length one. Note that $\rho_\text{jam}(+0)<\rho_*<\rho_\text{jam}(1)$.

\bibliography{references-packing}

\end{document}